 \documentclass[final,5p,times,twocolumn]{elsarticle}

\usepackage{amssymb}

\newcommand{\STO}{SrTiO$_3$}
\newcommand{\LAO}{LaAlO$_3$}

\newcommand{\MR}{magnetoresistance}

\newcommand{\etal}{\emph{et al.}}
\newcommand{\TDEG}{two dimensional electron gas}

\begin{document}

\begin{frontmatter}



\title{Anomalous magneto-transport at the superconducting interface between LaAlO$_3$ and SrTiO$_3$}


\author{M. Sachs}
\author{D. Rakhmilevitch}
\author{M. Ben Shalom}
\author{S. Shefler}
\author{A. Palevski}
\author{Y. Dagan}

\address{Raymond and Beverly Sackler School of Physics and Astronomy, Tel-Aviv University, 69978, Tel-Aviv, Israel}

\begin{abstract}
The magnetoresistance as a function of temperature and field for atomically flat interfaces between 8 unit cells of LaAlO$_3$ and SrTiO$_3$ is reported. Anomalous anisotropic behavior of the magnetoresistance is observed below 30 K for superconducting samples with carrier concentration of 3.5$\times$10$^{13}$ cm$^{-2}$. We associate this behavior to a magnetic order formed at the interface.
\end{abstract}


\end{frontmatter}

\par
It has been
shown that if \LAO~ is epitaxially grown on TiO$_2$ - terminated
\STO~ a \TDEG~ is formed at the interface between these insulators
\cite{OhtomoHwang}. This interface was latter shown to be
superconducting \cite{ReyrenSC} and magnetic \cite{BrinkmanSC}.
Recently Caviglia \etal~ have shown that the superconducting
transition temperature can be controlled by solely varying the
number of charge carriers at the interface using a gate
voltage.\cite{CavigiliaGating} These unexpected results and the
potential for development of high performance oxide based
electronics motivated an effort to understand the properties of
this interface and to improve it.
\par
Magnetic effects have been theoretically predicted for
\STO$\backslash$\LAO~ interfaces.\cite{Zhicheng_Zhong,
pentcheva:035112} Recent observations of magnetic hysteresis below
0.3K along with magneto-resistance oscillations with periodicity
proportional to $\sqrt{B}$ have been explained in terms of
commensurability of states formed at the terrace edges of the
\STO~ substrate.\cite{MOscillations}
\par
While superconductivity in this interface has been shown to be two
dimensional in nature \cite{ReyrenSC} the way such interface can
exhibit magnetic properties is still a puzzle.
\par
In this paper we
show that for carrier concentrations of $\sim3\times10^{13} cm^{-2}$
the \TDEG~ is superconducting at 100-300mK, yet, novel
magneto-transport effects are observed below 35K. Our data support
possible evidence for a magnetic order formed below this
temperature. A magnetic impurities scenario is ruled out.
\par
Eight unit cells of \LAO~ were deposited from a single crystal
target onto a TiO$_2$-terminated \STO~ by pulsed laser deposition.
We use pulse rate of 1Hz and energy
density of 1.5 J$\times$cm$^{-2}$ at oxygen pressure of 1$\times10^{-4}$ Torr and temperature
of $800^{\circ}C$ and a final annealing stage at a pressure of 200 mTorr O$_2$ and temperature of 400$^{\circ}C$ for 1 hour.
The deposition was monitored by reflection high
energy electron diffraction (RHEED). The maxima of the RHEED
intensity oscillations indicate a complete layer formation and
used as a measurement for the sample thickness. High
resolution transmission electron microscope imaging revealed a high
quality interface and confirmed the thickness measurement by the
RHEED. The
samples were patterned using reactive ion etch into Hall bars
with bridge dimensions of 50$\times$750 microns squared. Two bridges on the same substrate
were aligned 90$^\circ$ to each other (perpendicular or parallel to the terrace edges).
The resistance was measured in a dilution refrigerator using Lakeshore 370 AC resistance bridge and at liquid He temperatures using a standard 4 wire DC resistivity method.
\begin{figure}
\includegraphics[width=1\hsize]{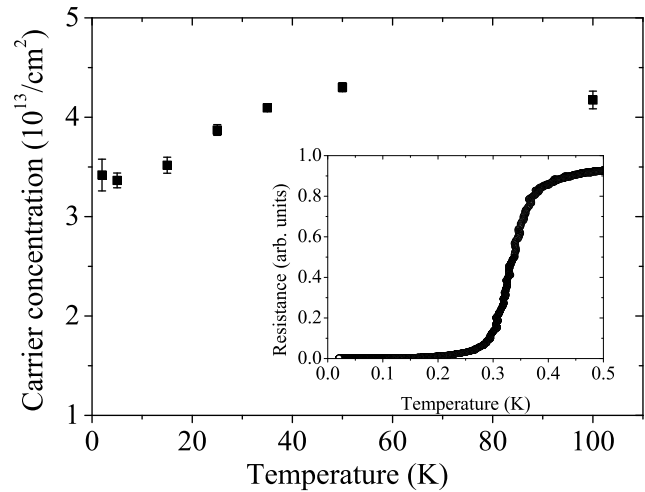}
\caption{(Color online) Number of charge carriers as inferred from the Hall resistivity  as a function of temperature. Note the small variation of the Hall number with temperature. Insert: resistance as a function of temperature below 0.5K .\label{Hall(T)}}
\end{figure}
\par
In Fig. ~\ref{Hall(T)} we present the number of charge carriers as a function of temperature for bridge $2$ and the normalized resistance for another bridge on the same film at low temperatures.
\par
\begin{figure}
\includegraphics[width=1\hsize]{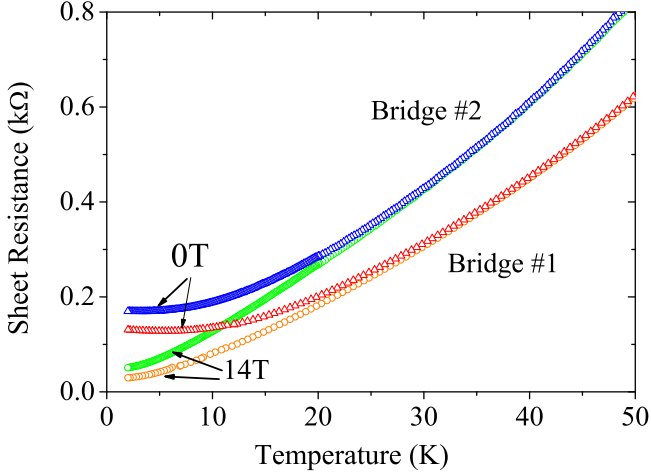}
\caption{(Color online) The sheet resistance for bridge $1$ and Bridge $2$ at zero field and at a magnetic field of 14 Tesla applied parallel to the current. The two curves merge above $\sim$30K).\label{RT_Hpar}}
\end{figure}

The magnetoresistance (MR) is strongly anisotropic both in plane and out of plane.\cite{benshalom} When the magnetic field is applied perpendicular to the interface a large positive MR is observed while
for field applied parallel to the current a strong negative MR is seen.
In Fig. ~\ref{RT_Hpar} The sheet resistance as a function of temperature is shown at zero field (circles) and at 14 Tesla applied parallel to the interface and to the current (triangles)
for bridges $1$ and $2$. First, we note that these bridges are rather similar
despite the fact that one of them is patterned along the terraces while the other is perpendicular to them. The small difference could be due to the time elapsed between the measurements of the two bridges (about 10 days).
The similarity between the bridges rules out the terraces as the origin for the strong magnetoresistance effects.
The strong parallel MR must therefore be related to a magnetic scattering. Above 30K the zero field curve and the 14 Tesla
one merge together. This suggests that the magnetic effects responsible for the parallel negative MR onset below 30K.
We have previously reported in-plane anisotropy of the parallel MR and related it to spin-orbit
coupling resulting in the anisotropic \MR, which is well known for magnetic materials.\cite{benshalom}
\par
\begin{figure}
\includegraphics[width=1\hsize]{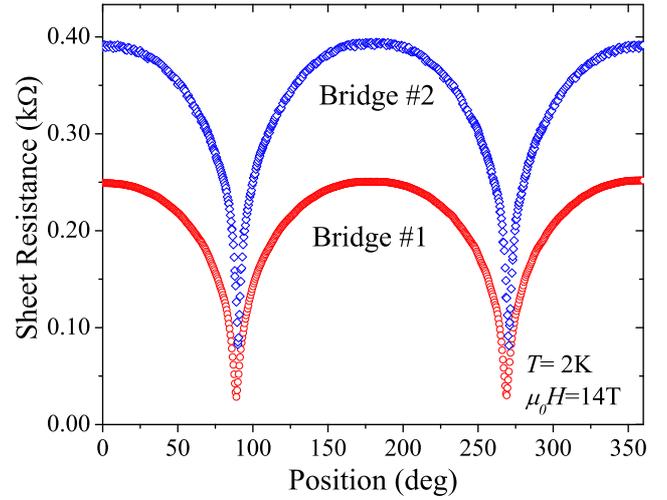}
\caption{The \MR~ as a function of the angle between the perpendicular
to the interface and the magnetic field.\label{rotation}}
\end{figure}

In Fig.~\ref{rotation} the MR for the two bridges at 14 Tesla is
plotted as a function of the angle $\varphi$ between the magnetic field and
the normal to the film. $90^\circ$
corresponds to in plane magnetic field applied parallel to the interface and to
the current.
The angular dependence of the MR shown in Fig.\ref{rotation} is extremely sharp
around $90^{\circ}$.
The relative MR defined as $\frac{R(H,\varphi)-R(H=0)}{R(H=0)}$ changes sign at $87^\circ$ (or $93^\circ$).

The fact that the MR changes sign for a variation of
$3^{\circ}$ implies that a small perpendicular field component is
sufficient to mute the mechanism responsible for the parallel
negative MR. This is due to the fact that when
$\varphi=93^{\circ}$ the parallel field component is almost
unchanged (13.98 Tesla) while the perpendicular component is only
0.73 Tesla. Such a component is too small to induce any orbital MR.\cite{benshalom}

The strong anisotropy of
the \MR~ is a key observation in our study. The only element in our system with such strong
directionality is the interface itself. This gives possible evidence for the
existence of magnetic order confined to a few layers near the
interface. This magnetic order vanishes above 30K
according to the data in Fig.\ref{RT_Hpar} (for the carrier density
and \LAO~ thickness under study).
We also note that the MR effects become more pronounced at lower temperatures
where superconductivity appears (not shown, to be published).
\par
In summary, below 30K a magnetic phase emerges.
This phase is extremely sensitive to an out of plane magnetic field. This sensitivity is
unclear to us, yet, it rules out magnetic impurities as the origin
for this effect. Impurity effects are isotropic, in strong contrast with our data.
Our data therefore suggest that the magnetic order appearing below 30K is confined to
the vicinity of the interface.
For carrier density and thickness under study superconductivity is seen together with the magnetic effects.
Since superconductivity disappears at rather low fields and the effects reported here are observed at stronger magnetic fields it is not clear if the magnetic effects exist at zero field. Further experiments are needed to clarify this question.
\par
This research was partially supported by the Bikura program of the Israeli Science Foundation Agreement No. 1543/08 by the Israel Science Foundation
Agreement No. 1421/08 and by the Wolfson Family Charitable Trust.
A portion of this work was performed at the National High Magnetic Field
Laboratory, which is supported by NSF Cooperative Agreement No.
DMR-0654118, by the State of Florida, and by the DOE.




\end{document}